\RequirePackage[hyphens]{url}
\documentclass[letterpaper,twocolumn,10pt]{article}

\usepackage{usenix}
\usepackage{tikz}
\usetikzlibrary{tikzmark}

\makeatletter
\newlength{\mylength}

\newsavebox{\mybox}

\makeatother

\usepackage[normalem]{ulem}

\usepackage{verbatimbox}

\usepackage{amsmath}
\usepackage{amsthm}
\usepackage{amssymb}
\usepackage{booktabs}
\usepackage{caption}
\usepackage{color}
\usepackage{graphicx}
\usepackage{listings}
\usepackage{times}
\usepackage{comment}

\newcommand{\systemname}{\textsc{Scalene}}

\title{Triangulating Python Performance Issues with \systemname{}}

\author{
  {\rm Emery D. Berger} \\
  College of Information and Computer Sciences \\
  University of Massachusetts Amherst \\
  \texttt{emery@cs.umass.edu}\\
  \and
      {\rm Sam Stern} \\
      College of Information and Computer Sciences \\
      University of Massachusetts Amherst \\
      \texttt{jstern@cs.umass.edu} \\
      \and
          {\rm Juan Altmayer Pizzorno} \\
          College of Information and Computer Sciences \\
          University of Massachusetts Amherst \\
          \texttt{jpizzorno@cs.umass.edu}\\
}

\newcommand{\punt}[1]{}

\punt{
\microtypesetup{
  final,
  tracking=true,
  kerning=true,
  spacing=true,
  factor=1100,
  stretch=20,
  shrink=20
}
\SetTracking{encoding={*}, shape=sc}{0} 
}

\definecolor{mygreen}{rgb}{0,0.6,0}
\definecolor{mygray}{rgb}{0.5,0.5,0.5}
\definecolor{mymauve}{rgb}{0.58,0,0.82}
\usepackage{inconsolata}

\begin{document}

  \maketitle
    
  \begin{abstract}
    \vspace{0.5em}
    This paper proposes \systemname{}, a profiler specialized for
Python. \systemname{} combines a suite of innovations to precisely and
simultaneously profile CPU, memory, and GPU usage, all with low
overhead. \systemname{}'s CPU and memory profilers help Python
programmers direct their optimization efforts by distinguishing
between inefficient Python and efficient native execution time and
memory usage. \systemname{}'s memory profiler employs a novel sampling
algorithm that lets it operate with low overhead yet high
precision. It also incorporates a novel algorithm that automatically
pinpoints memory leaks, whether within Python or across the
Python-native boundary. \systemname{} tracks a new metric called copy
volume, which highlights costly copying operations that can occur when
Python silently converts between C and Python data representations, or
between CPU and GPU. Since its introduction, \systemname{} has been
widely adopted, with over 500,000 downloads to date. We present
experience reports from developers who used \systemname{} to achieve
significant performance improvements and memory savings.

  \end{abstract}

  \section{Introduction}
  Python is now one firmly established as one of the most popular
programming languages, with first place rankings from
TIOBE~\cite{tiobe-index} and IEEE Spectrum~\cite{ieeeplrank2022},
second place on the Redmonk Rankings~\cite{redmonk-rankings}, and
fourth place in the 2022 Stack Overflow Developer
Survey~\cite{stack-overflow-2022-survey}. Large-scale industrial users
of Python include Dropbox~\cite{python-at-dropbox},
Facebook~\cite{python-at-facebook},
Instagram~\cite{python-at-instagram},
Netflix~\cite{python-at-netflix}, Spotify~\cite{python-at-spotify},
and YouTube~\cite{python-at-youtube}.

\begin{figure*}[!t]
\centering
\includegraphics[width=\linewidth]{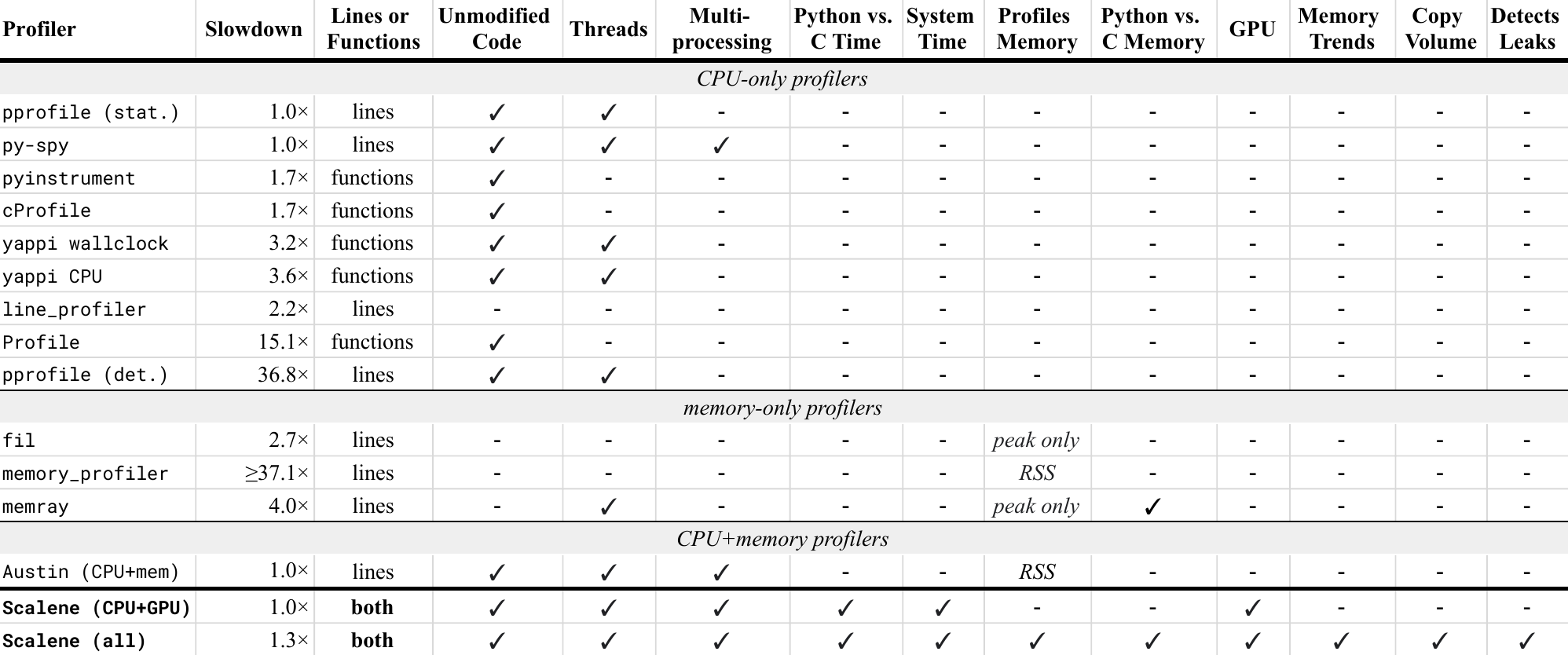}
\vspace{1em}
\caption{
        \textbf{\systemname{} vs. past Python
 profilers.}  \systemname{} provides vastly more information than past
 Python profilers, with more accurate memory profiling ($\S$\ref{sec:memory-profiling-accuracy}) and with
 low overhead ($\S$\ref{sec:cpu-profiling-overhead}, $\S$\ref{sec:memory-profiling-overhead}). Most past profilers
 ($\S$\ref{sec:other-profilers}) exclusively profile either CPU time
 or memory; \systemname{} simultaneously profiles CPU, GPU, and
 (optionally) memory, and comprises a suite of unique features backed
 by novel algorithms.  \label{fig:profiler-comparison} }
\end{figure*}

At the same time, Python is (in)famously slow. The standard Python
implementation, known as CPython, is a stack-based bytecode
interpreter written in C~\cite{enwiki:1095361531}.
Pure Python code typically runs 1--2 orders
of magnitude slower than native code. As an extreme example, the
Python implementation of matrix-matrix multiplication takes more than
$60,000\times$ as long as the native BLAS version.

Python's performance costs are nearly matched by its high memory
overhead. Python data types consume dramatically more memory than
their native counterparts. For example, the integer 1 consumes 4
bytes in C, but 28 bytes in Python; \texttt{"a"}
consumes 2 bytes in C, but 50 bytes in Python.
This increased space demand is primarily due to metadata that Python
maintains for every object, including reference counts and dynamic
type information. Python is also a garbage collected language; because
garbage collection delays memory reclamation, it can further increase
the amount of memory consumed compared to native code~\cite{1094836}.

Because of these costs, one of the most effective ways for Python
programmers to optimize their code is to identify performance-critical
and/or memory-intensive code that uses pure Python, and replace it with
native libraries. Python's ecosystem includes
numerous high-performance packages with native implementations, which
are arguably the key driver of its adoption and popularity. These
libraries include the NumPy numeric library~\cite{oliphant2006guide},
the machine learning libraries SciKit-Learn~\cite{pedregosa2011scikit}
and TensorFlow~\cite{tensorflow2015-whitepaper,abadi2016tensorflow},
among many others. By writing code that makes effective use of these
packages, Python programmers can sidestep Python's space and time
costs, and at the same time take full advantage of underlying hardware
resources like multiple cores, vector instructions, and GPUs.

Unfortunately, past Python profilers---which can be viewed as ports of
traditional profilers for native code---fail to meet this
challenge. While the approaches they embody are satisfactory for
profiling native code, they fall short in the context of Python. We
contend that Python programmers need a profiler that provides a
holistic, granular view of their program's execution to help them
identify and remedy inefficiencies in their programs, especially
in steering them towards and improving the efficiency of their use of
native libraries.

This paper proposes \systemname{}
, a profiler comprising a suite of profiling
innovations designed specifically for Python. Unlike all past Python
profilers, \systemname{} simultaneously profiles CPU, memory usage,
and GPU usage. It provides fine-grained information targeted
specifically at the problems of optimizing Python code. In
particular, \systemname{} teases apart time and memory consumption
that stems from Python vs. native code, revealing where they can
optimize by switching to native code. \systemname{} reports a new
metric, \emph{copy volume}, that helps identify costly (and often
inadvertent) copying across the Python/native divide, or copying
between CPU and GPU. Its memory profiler accurately tracks memory
consumption over time, and automatically identifies memory leaks,
whether within Python or spanning the Python-native code divide. Its
GPU profiler tracks GPU utilization and memory consumption, letting it
identify when native libraries are not being used to their best
advantage. At the same time, \systemname{} imposes low overhead
(median: 0\% for CPU+GPU, 32\% for CPU+GPU+memory).

Since its introduction,
\systemname{} has become a popular tool among Python developers, with
over 500,000 downloads to date. We report on case studies supplied by
external users of \systemname{}, including professional Python open
source developers and industrial users, highlighting how \systemname{}
helped them diagnose and then remedy their performance problems,
leading to improvements ranging from 45\% to $125\times$.

This paper makes the following contributions: it
proposes \textbf{\systemname{}}, a profiler specifically tailored to
Python; it presents several novel algorithms, including (1) its
algorithm for attributing time consumption to Python or
native code; (2) its sampling-based memory profiling that is
both accurate and low overhead; and (3) its automatic memory leak
detector, which identifies leaks with low overhead. It also introduces
and demonstrates the value of a new metric, \emph{copy volume}, that
surfaces hidden costs due to copying.

\begin{figure*}[!t]
\centering
\includegraphics[width=\linewidth]{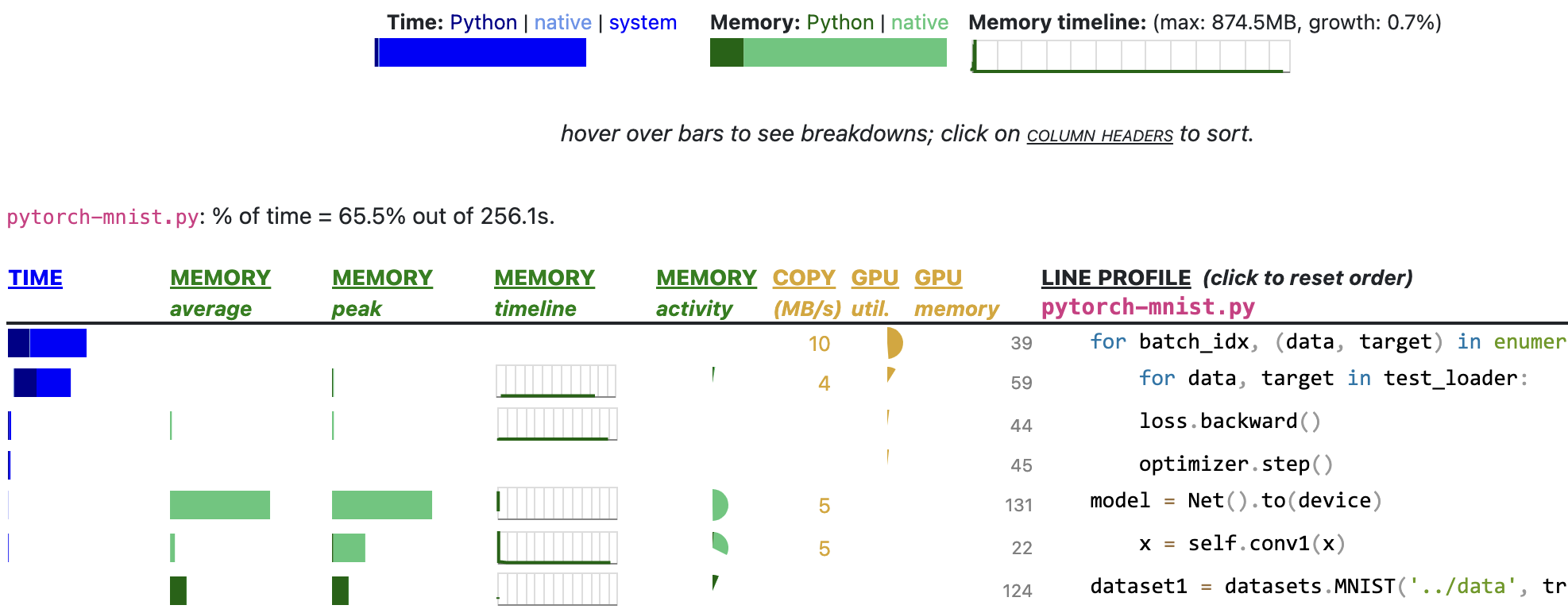}
\vspace{1em}
\caption{
        \textbf{An example profile from \systemname{}'s web UI},
        sorted in descending order by GPU utilization. The top graphs
        provide a summary for the entire program, with more detailed
        data reported for each active line (and, not shown, for each
        function). CPU time is in blue, with different shades
        reflecting time taken in Python code, native code, or
        system/GPU time ($\S$\ref{sec:cpu-implementation}).  Average
        and peak memory consumption is in green, with different shades
        distinguishing memory consumed by Python objects vs. native
        ones ($\S$\ref{sec:memory-implementation}); the memory timeline
        depicts memory consumption over time ($\S$\ref{sec:gui}). Copy
        volume is in yellow ($\S$\ref{sec:copy-volume}), as well as GPU
        utilization and GPU memory consumption
        ($\S$\ref{sec:gpu-implementation}). Hovering over bars
        provides detailed statistics in hovertips.\label{fig:profile-example}
        }
\end{figure*}

  \label{sec:implementation}

The next sections explain \systemname{}'s implementation and
algorithms. We first outline how \systemname{} efficiently performs
line-level CPU profiling, focusing on its approach to teasing apart
time spent running in the Python interpreter from native code
execution and system time ($\S$\ref{sec:cpu-implementation}). We then
describe \systemname{}'s memory profiling component
($\S$\ref{sec:memory-implementation}), including
its \emph{threshold-based} sampling approach that reduces overhead
while ensuring accuracy, its memory leak detection algorithm,
and how it tracks copy
volume.  We then explain
how \systemname{} profiles GPU utilization and memory consumption
($\S$\ref{sec:gpu-implementation}). Finally, we present technical
details underpinning \systemname{}'s user interface
($\S$\ref{sec:gui}). We then present our evaluation
($\S$\ref{sec:evaluation}) and a number of case studies of user
experiences with \systemname{} ($\S$\ref{sec:case-studies});
we conclude with a discussion of related work ($\S$\ref{sec:related_work}).

\section{CPU Profiling}
\label{sec:cpu-implementation}


\label{sec:python-versus-c}

\systemname{}'s CPU profiler employs sampling, but unlike past
profilers, it leverages how Python delivers signals to extract more
granular information. As we explain, Python signals complicate this
task.

Sampling profilers like \systemname{} work by periodically
interrupting program execution and examining the current program
counter. Given a sufficiently large number of samples, the number of
samples each program counter receives is proportional to the amount of
time that the program was executing. Sampling can be triggered by the
passage of real (wall-clock) time, which accounts for CPU time as well
as time spent waiting for I/O or other events, or virtual time (the
time the application was scheduled for execution), which only accounts
for CPU time.

Unfortunately, in Python, using sampling to drive profiling can lead
to erroneous profiles.  Like other scripting languages such as Perl
and Ruby, Python only delivers signals to the main
thread~\cite{signals-and-threads}. Also like those languages, Python
defers signal delivery until the virtual machine (i.e., the
interpreter loop) regains control, and only checks for pending signals
after specific opcodes such as jumps.

The result is that, during the entire time that Python spends
executing external library calls, no timer signals are delivered.  The
effect can be that the profiler will reflect \emph{no time} spent
executing native code, no matter how long it actually took. In
addition, because only main threads are interrupted, sampling
profilers can fail to account for any time spent in child threads

In fact, this is a failure mode for one of the profilers we examine
here. \texttt{pprofile (stat.)} relies exclusively on timer signal
delivery to perform CPU profiling.  Because it fails to cope with the
cases described above, this profiler reports zero elapsed time for all native
execution or code executing in multiple threads.

\begin{figure}[!t]
\centering
  \includegraphics[width=0.8\linewidth]{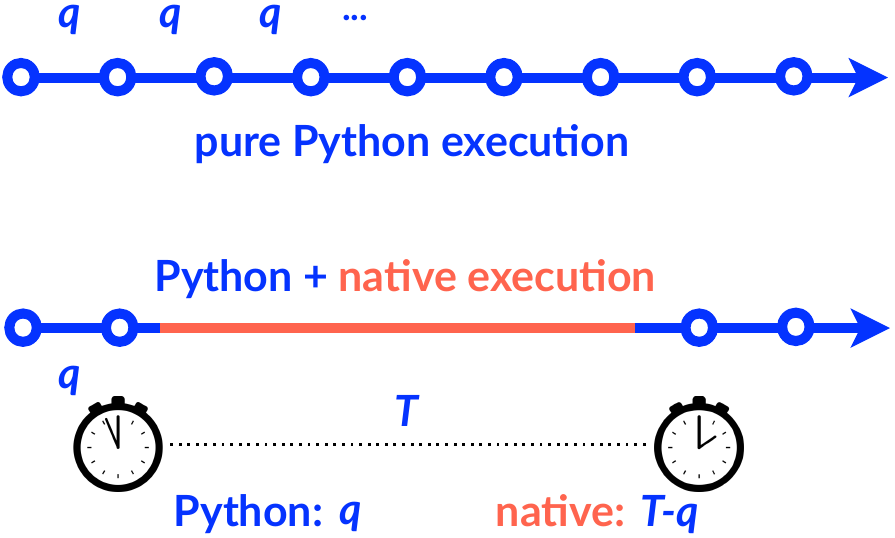}
  \caption{\textbf{Overview of \systemname{}'s inference of Python vs. native execution.} Sampling profilers depend on regular timer interrupts, but Python defers all signals when running native code, leading to the appearance of no time spent executing that code. \systemname{} leverages this apparent limitation to accurately attribute time spent executing Python and native code ($\S$\ref{sec:inferring-native-time}) in the main thread; it uses a different algorithm for code running in threads ($\S$\ref{sec:cpu-profiling-threads}).\label{fig:interrupt-diagram}}
  \vspace{1em}
\end{figure}

\subsection{Accurate Python-C Profiling}
\label{sec:inferring-native-time}

\systemname{}'s CPU profiler turns these limitations of Python signals to its
advantage, inferring whether a line spent its time executing Python or native (C) code.
It leverages the following insight: \emph{any delay in
signal delivery corresponds to time spent executing outside the
interpreter}. That is, if \systemname{}'s signal handler received the
signal immediately (that is, in the requested timing interval), then
all that time must have been spent in the interpreter. If it was
delayed, it must be due to running code outside the interpreter, which
is the only cause of delays (at least, in virtual time).

Figure~\ref{fig:interrupt-diagram} depicts how \systemname{} handles
signals and attributes time to either Python or native code.
\systemname{} tracks time between interrupts
recording the current virtual time whenever it receives a CPU timer
interrupt (using \texttt{time.process\_time()}). When it receives the
next interrupt, it computes $T$, the elapsed virtual time, and compares
it to the timing interval $q$ (for quantum).

\systemname{} uses these values to attribute time spent to Python or native code.
Whenever \systemname{} receives a signal, \systemname{} walks the
Python stack until it reaches code being profiled (that is, outside of
libraries or the Python interpreter itself), and attributes time to
the identified line of code. \systemname{} maintains two counters for
every line of code being profiled: one for Python, and one for C
(native) code. Each time a line is interrupted by a
signal, \systemname{} increments the Python counter by $q$, the timing
interval, and it increments the C counter by $T-q$, the delay.


\subsection{Accurate Python-C Profiling of Threads}
\label{sec:cpu-profiling-threads}

The approach described above attributes execution time for Python
vs. C code in the main thread, but it does not attribute execution
time at all for subthreads, which, as described above, never receive
signals. To attribute Python and C time for code running in
subthreads, \systemname{} applies a different algorithm, this time
leveraging a combination of Python features: \emph{monkey
patching}, \emph{thread enumeration}, \emph{stack inspection},
and \emph{bytecode disassembly}.


Monkey patching refers to the redefinition of functions at
runtime. \systemname{} uses monkey patching to ensure that signals are
always received by the main thread, even when the main thread is
blocking (e.g., waiting to join with child threads). \systemname{}
replaces blocking functions like \texttt{threading.join} with ones
that always use timeouts. It sets these timeouts to Python's own
thread quantum, obtained
via \texttt{sys.getswitchinterval()}. Replacing these calls ensures
that the main thread yields periodically, allowing signals to be
delivered.

In addition, to attribute execution times correctly, \systemname{}
maintains a status flag for every thread, all
initially \emph{executing}. In each of the calls it intercepts,
before \systemname{} issues the blocking call, it sets the
calling thread's status as \emph{sleeping}.  Once that thread returns
(either after successfully acquiring the desired resource or after a
timeout), \systemname{} resets the status of the calling thread to
executing. \systemname{} only attributes time to currently executing
threads.

Now, when the main thread receives a signal, \systemname{}
invokes \texttt{threading.enumerate()} to collect a list of all
running threads. It then obtains the Python stack frame from each
thread using Python's \texttt{sys.\_current\_frames()} method. As
above, \systemname{} walks the stack to find the appropriate line of
code to attribute execution time.

Finally, \systemname{} uses bytecode disassembly (via the \texttt{dis}
module) to distinguish between time spent in Python vs. C
code. Whenever Python invokes an external function, it does so using a
bytecode whose textual representation is either \texttt{CALL\_FUNCTION},
\texttt{CALL\_METHOD}, or, as of Python 3.11, \texttt{CALL}.
\systemname{} builds a map of all such bytecodes at startup.

For each running thread, \systemname{} checks the stack and its
associated map to determine if the currently executing bytecode is a
call instruction. \systemname{} can use this
information to infer with high likelihood whether the thread is
currently executing Python or C code.

If a thread is running Python code, it is likely to spend almost no
time in a bytecode before executing another Python bytecode. By
contrast, if if it is running C code, it will be ``stuck'' on
the \texttt{CALL} bytecode for the duration of native
execution. Leveraging this lets \systemname{} accurately attribute
execution time straightforwardly: if it finds that a stack is
executing \texttt{CALL}, \systemname{} assigns time elapsed to the C
counter; otherwise, it assigns time elapsed to the Python counter.

\section{Memory and Copy Volume Profiling}
\label{sec:memory-implementation}

Almost all past profilers either report CPU time or memory
consumption; \systemname{} reports both, at a line granularity.  It is
vital that \systemname{} track memory both inside Python and out, as
external libraries are often responsible for a considerable fraction
of memory consumption.

\subsection{Intercepting Allocation Calls}

\systemname{} intercepts all system allocator calls
(\texttt{malloc}, \texttt{free}, etc.) as well as Python internal
memory allocator by inserting its own ``shim'' memory allocator, using
Python's built-in memory hooks. This two-fold approach lets \systemname{} distinguish between native
memory allocated by libraries and Python memory allocated in the
interpreter.

The shim allocator extends and uses code from the
Heap Layers memory allocator infrastructure~\cite{DBLP:conf/pldi/BergerZM01}; \systemname{} injects it
via library interposition before Python begins executing 
using \texttt{LD\_PRELOAD} on Linux and \texttt{DYLD\_INSERT\_LIBRARIES} on Mac OS X.
To interpose on Python's internal memory allocator, \systemname{} uses Python's custom allocator API
(\texttt{PyMem\_SetAllocator}).


Each shim allocator function handles calls by sampling for inclusion in the profiling
statistics ($\S$\ref{sec:memory-sampling}) and then passing these to the original (Python or
system) allocator.
A complication arises from the fact that the Python allocators themselves may handle allocation
requests by calling into the system allocator.
To avoid counting Python allocations also as native allocations, \systemname{} sets a flag,
stored in thread-specific data, indicating it is within a memory allocator.
When a shim allocator function is called with this flag set, it skips over the profiling,
just forwarding to the original allocator.
This approach both avoids double counting and simplifies writing profiling code, as it can allocate
memory normally without causing infinite recursion.

\begin{figure}[!t]
\centering
  \includegraphics[width=\linewidth]{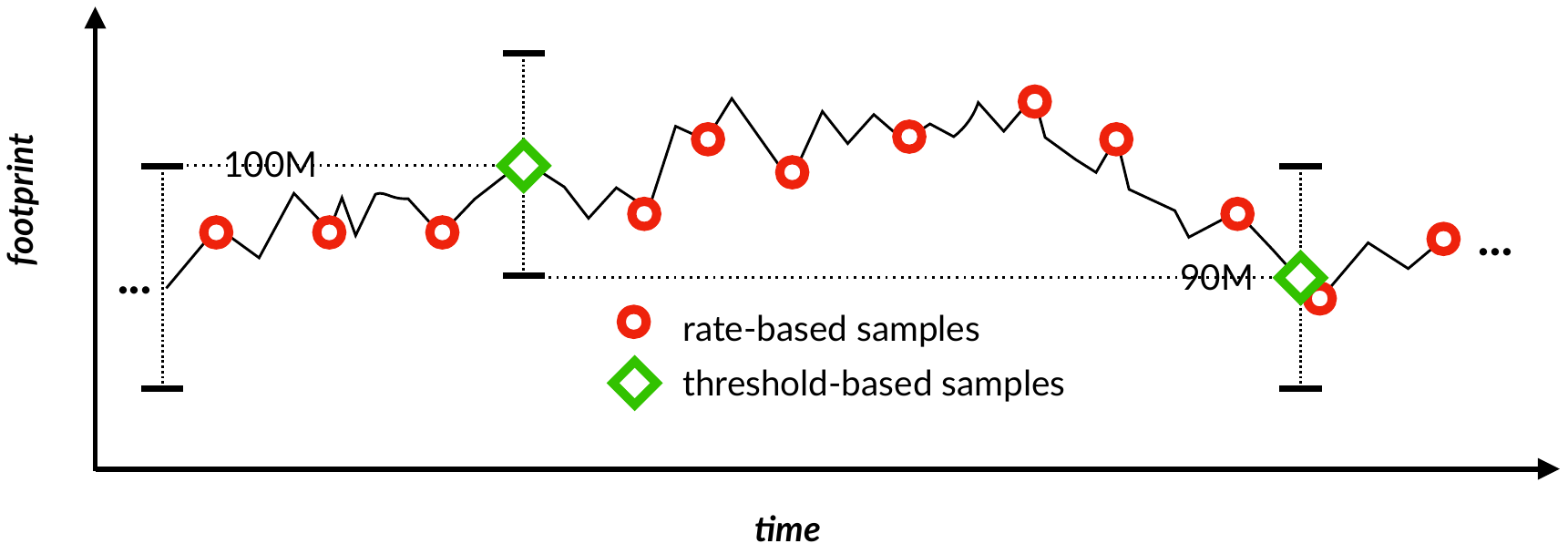}
  \vspace{1em}
  \caption{\textbf{Threshold-Based vs. Rate-Based Sampling.} \systemname{} employs a novel sampling scheme that only triggers when memory use grows or declines beyond a set threshold, letting it capture all significant changes in footprint (beyond a given granularity, here 10MB) with low overhead. ($\S$\ref{sec:memory-sampling}).\label{fig:memory-sampling-diagram}}
  \vspace{1em}
\end{figure}

\subsection{Threshold-Based Sampling}
\label{sec:memory-sampling}

The standard approach to sampling memory profilers, as exemplified by
several non-Python memory profilers in Android, Chrome, Go, and
Google's \texttt{tcmalloc}~\cite{chrome-memory-sampling} and in Java TLAB
based sampling~\cite{tlab-profiling}, use a \emph{rate-based} sampling
approach. This sampler triggers samples at a rate proportional to the
number of bytes allocated or freed. In effect, each byte allocated or
freed corresponds to a Bernoulli trial with a given probability $p$ of
sampling; e.g., if $p = 1/\texttt{T}$, then (in expectation) there
will be one sample per \texttt{T} bytes. In practice, for efficiency,
these samplers initialize counters to random numbers drawn from a
Poisson process or a geometric distribution with the same
parameter. Each allocation and free then decrements this counter by
the number of bytes allocated and freed, and triggers a sample when
the counter drops below 0.

By contrast, the \systemname{} sampler introduces
\emph{threshold-based} sampling.  The allocator maintains a count of
all memory allocations and frees, in bytes. Once the absolute
difference between allocations and frees crosses a threshold ($|A - F|
>= T$), \systemname{} triggers a sample, corresponding to appending an
entry to a sampling file and resets the counters.
Figure~\ref{fig:memory-sampling-diagram} illustrates this operation.
The sampling threshold $T$ is currently set to a prime number slightly
above 10MB; \systemname{} uses a prime number to reduce the risk of
stride behavior interfering with sampling.

Threshold-based sampling has several advantages over rate-based
sampling. Unlike rate-based sampling, which is triggered by all
allocation activity (even when it has almost no effect on footprint),
threshold-based sampling is only triggered by significant memory use
growth or decline.  Table~\ref{tab:sampling-comparison} shows the
dramatic reduction in the number of samples, as high as $676\times$ (median: $18\times$)
fewer. This reduced number of samples translates directly to lower
runtime overhead.

At the same time, threshold-based sampling deterministically triggers
a new sample whenever a significant change in footprint occurs. This
approach improves repeatability over rate-based sampling (which is
probabilistic) and avoids the risk of missing these changes.

Crucially, threshold-based sampling avoids two sources of bias
inherent to rate-based sampling.  Rate-based sampling can overstate
the importance of allocations that do not contribute to an increased
footprint since it does not take memory reclamation or footprint into
account. It also biases the attribution of memory consumption to lines
of code running Python code that exercises the allocator, rather than
code responsible for footprint changes.  By constrast, threshold-based
sampling filters out the vast number of short-lived objects that are
created by the Python interpreter itself, and only triggers based on
events that change footprint.



\subsection{Collecting and Processing Samples}

When a memory sample is taken, \systemname{} temporarily enables tracing
using Python's
\texttt{PyEval\_SetTrace}. Tracing remains active only until it detects execution has moved on from that line.
This approach lets \systemname{} properly account for average memory consumption per line.

Each entry in \systemname{}'s sampling file includes information about
allocations or frees, the fraction of Python (vs. native) allocations
in the total sample, as well as an attribution to a line of Python
source code.

\systemname{} attributes each sample to Python source code at the time the sample is taken.
It does so by obtaining the current thread's call stack from the interpreter and skipping over frames
until one within profiled source code is found.
This attribution needs to happen whenever a sample is taken, so it is implemented as a
C++ extension module, using read-only accesses to Python structures.
\systemname{} loads this module upon startup, which in turn uses a symbol exported by the
shim library to complete the linkage, making itself available to the shim.

A background thread in \systemname{}'s Python code reads from the sampling file and updates
the profiling statistics.
\systemname{} also tracks the
current memory footprint, which it uses both to report maximum memory
consumption and memory trends. \systemname{} records a timestamp and
the current footprint at each threshold crossing, which \systemname{} uses to
generate memory trend visualizations ($\S$\ref{sec:gui}).


\subsection{Memory Leak Detection}
\label{sec:memory-leak-detector}

Like other garbage-collected languages, Python can suffer from memory
leaks when references to objects are accidentally retained so that the
garbage collector cannot reclaim them. As in other garbage collected languages,
identifying leaks in Python programs is generally a slow, manual process.

In Python, the standard approach to identifying leaks is to first
activate \texttt{tracemalloc}, which records the size, allocation
site, and stack frame for each allocated object.  The programmer then
inserts calls at the appropriate place to produce a series of heap
snapshots, and then manually inspects snapshot diffs to identify
growing objects. This approach suffers from several drawbacks. First,
it is laborious and depends on a post hoc analysis of the
heap. Second, it can be quite slow. In our tests, just
activating \texttt{tracemalloc} can slow Python applications down by
$4\times$.


Instead, \systemname{} incorporates a novel sampling-based memory leak
detection algorithm that is both simple and efficient.  The algorithm
piggybacks on threshold-based sampling ($\S$\ref{sec:memory-sampling}).
Whenever the threshold-based sampler triggers because of memory
growth, \systemname{} checks to see if this growth has led to a new
high-water mark (maximum footprint). If so, \systemname{} records the
sampled allocation. Every call to \texttt{free} then checks to see
whether this object is ever reclaimed. This checking is cheap and
highly predictable, consisting of a pointer comparison that is almost always
false.

\paragraph{Leak Score:}
At the next crossing of a maximum, \systemname{} updates a \emph{leak
score} for the sampled object. The leak score tracks the historic
likelihood of reclamation of the sampled object, and consists of a
pair of $(\texttt{frees}, \texttt{mallocs})$. \systemname{} first
increments the \texttt{mallocs} field when it starts tracking an
object, and then increments the \texttt{frees} field only if it
reclaimed the allocated object. It then resumes tracking with a newly
sampled object.

Intuitively, leak scores capture the likelihood that an allocation
site is leaking. A site with a high number
of \texttt{mallocs} and no \texttt{frees} is a plausible leak. By
contrast, a site with a matching number of \texttt{mallocs}
and \texttt{frees} is probably not a leak.  The more observations we
make, the higher the likelihood that we are observing or ruling out a
leak.

We use Laplace's Rule of Succession to compute the likelihood of a
success or failure in the next Bernoulli trial, given a history of
successes and failures~\cite{zabell1989rule}. Here, successes
correspond to reclamations (\texttt{frees}) and failures are
non-reclamations (\texttt{mallocs} - \texttt{frees}). According to the
Rule of Succession,
\systemname{} computes the leak probability
as $1.0 - (\texttt{frees} + 1) / (\texttt{mallocs} -
\texttt{frees} + 2)$.

\paragraph{Leak Report Filtering and Prioritization:}

To provide maximal assistance to Python developers, \systemname{}
filters and augments leak information. First, to limit the number of
leak reports, \systemname{} only reports leaks whose likelihood
exceeds a 95\% threshold, and only when the slope of overall memory
growth is at least 1\%. Second, \systemname{} makes it possible for
developers to prioritize leaks by associating each leak with an
estimated \emph{leak rate}. Leak rate corresponds to the average
amount of memory allocated at a given line divided by time elapsed, in
MB per second. We expect Python programmers to focus their
attention on high-confidence leaks with a high leak rate, since these
are the most serious.

\subsection{Copy Volume}
\label{sec:copy-volume}

\systemname{} uses sampling to collect information about \emph{copy volume}
(megabytes per second of copying) by line. This metric,
which \systemname{} introduces, helps identify costly (and often
inadvertent) copying across the Python/native divide, or copying
between CPU and GPU.

The \systemname{} shim
library used for memory allocation also interposes on \texttt{memcpy},
which is invoked both for general copying (including to and from the
GPU, and copying across the Python/C boundary). As with memory
allocations,
\systemname{} writes an entry to a sampling file once a threshold
number of bytes has been copied. However, unlike memory sampling, copy volume
sampling employs classical rate-based sampling. The
current \texttt{memcpy} sampling rate is set at a multiple of the
allocation sampling rate.

\section{GPU Profiling}
\label{sec:gpu-implementation}

\systemname{} performs both line-granularity GPU utilization and memory profiling on
systems equipped with NVIDIA GPUs. This feature helps Python programmers identify
whether they are efficiently making use of their GPUs.

\systemname{} piggybacks GPU sampling on top of its CPU sampler. Every time
\systemname{} obtains a CPU sample,
it also collects the total currently used GPU memory and utilization,
which it associates with the currently executing line of code. Whenever
possible, it employs per-process ID accounting, which can substantially
increase accuracy in a shared GPU setting.

At startup, \systemname{} checks to see if per-process ID accounting
has been enabled on the attached NVIDIA GPU. If not, \systemname{}
offers to enable it, a process that requires that the user
invoke \systemname{} once with super-user privileges.


\section{GUI Design and Implementation}
\label{sec:gui}

\systemname{}'s primary user interface is web-based, though it also offers
a non-interactive rich text-based CLI. The web UI, written in
JavaScript, uses Vega-Lite to generate its visualizations, including
bar graphs, pie charts, and the line graphs used for visualizing
memory consumption over time~\cite{2017-vega-lite}. To avoid CORS
issues, \systemname{} produces a single HTML payload that includes the
actual JSON-based profile (which it also outputs as a separate file),
and then launches a local browser tab to open it. This approach also
makes it trivial to upload, share, or archive profiles.

In the UI, \systemname{} not only reports net memory consumption per
line, but also reports memory usage over time, both for the program as
a whole and for each individual line. Figure~\ref{fig:profile-example}
presents several examples. The x-axis corresponds to execution time, and
the y-axis corresponds to the footprint of the program, as seen by
that line of code.

Because it can be expensive to visualize graphs with large numbers of
points, \systemname{} limits the number of points it outputs in its
JSON payload and HTML output. Prior to generating the profile
output, \systemname{} applies the Ramer-Douglas-Peucker (RDP)
algorithm~\cite{douglas1973algorithms,ramer1972iterative} to each
line's memory footprint log (if any). The RDP algorithm aims to reduce
the total number of points while preserving the overall shape of the
curve. The RDP algorithm depends on a parameter $\epsilon$ which
corresponds to a distance parameter below which RDP merges adjacent points;
\systemname{} sets $\epsilon$ to a value that approximately reduces the total number of points
to a manageable size (100 points). Sometimes this process fails to
reduce the number of points sufficiently. To guarantee that the number
of points is always bounded, after applying RDP, \systemname{}
randomly downsamples all memory logs to exactly 100 points.

To further ensure the scalability of the user interface, \systemname{}
only reports lines of code that are responsible for at least 1\% of
execution time (CPU or GPU) or at least 1\% of total memory
consumption, along with the preceding and following line. This
approach guarantees that a \systemname{} profile never contains more
than 300 lines. In practice, profiles are generally skewed and
resulting profilers are often far more abbreviated.


  \section{Evaluation}
  \label{sec:evaluation}

Our evaluation answers the following questions: (1) How
does \systemname{}'s CPU profiling accuracy compare to other CPU
profilers? ($\S$\ref{sec:cpu-profiling-accuracy}) (2) How
does \systemname{}'s memory profiling accuracy compare to other memory
profilers? ($\S$\ref{sec:memory-profiling-accuracy}) (3) How
does \systemname{}'s CPU profiling overhead compare to other CPU
profilers? ($\S$\ref{sec:cpu-profiling-overhead}) (4) How
does \systemname{}'s memory profiling overhead compare to other memory
profilers? ($\S$\ref{sec:memory-profiling-overhead})

\subsection{Experimental Setup}

Our prototype of \systemname{} consists of roughly 3,500 lines of
Python 3 code and 1,700 lines of C++-17 code; its user interface
comprises 800 lines of JavaScript, excluding white space and
comments as measured by \texttt{cloc}~\cite{adanial_cloc}. This
prototype runs on Linux, Microsoft Windows, and Mac OS X, for Python
versions 3.8 and higher; we report Linux results here. We use the
latest version of \systemname{}, released 12/08/2022.

We perform all experiments on an 8-core 4674 MHz AMD Ryzen 7, equipped
with 32GB of RAM and an NVIDIA GeForce RTX 2070 GPU, running Linux
5.13.0-35-generic. All C/C++ code is compiled with g++ version 9,
and we use CPython version 3.10.9 (release date 12/06/2022)
For overhead numbers, we report the interquartile
mean of 10 runs.

\subsection{CPU Profiling Accuracy}
\label{sec:cpu-profiling-accuracy}

Here, we explore a specific threat to the accuracy of
Python CPU profilers. We hypothesize that some might exhibit
a \emph{probe effect} that could distort the time spent by
applications. Specifically, we suspect that Python profilers that
rely on Python's tracing facility might exhibit a bias caused by
tracing triggering both on function calls and lines of code,
dilating the apparent time spent in function calls. We call this
phenomenon \emph{function bias}; we hypothesize that
sampling-based profilers like \systemname{} would not suffer from
this bias.

We wrote a microbenchmark to test this hypothesis. The microbenchmark
executes a varying number of iterations of two semantically identical
functions: one invokes another function inside its loop, while the
other inlines the same logic. Our experiments where vary a parameter
of the microbenchmark---the amount of time spent in one function
versus the other---and compare the profiler results to the ground
truth, as measured with high resolution timers.

\punt{
\begin{footnotesize}
\begin{lstlisting}[language=python, basicstyle=\ttfamily\footnotesize, numbers=none]
def main():
    x = 0
    x = inline_loop(x)
    x = fn_call_loop(x)
    
def inline_loop(x):
    for i in range(NUM_INLINE):
        x = x | (x >> 2) | (i & x)
    return x

def fn_call_loop(x):
    for i in range(NUM_FN_CALLS):
        x = x | do_work_fn(x, i)
    return x

def do_work_fn(x, i):
    return (x >> 2) | (i & x)
\end{lstlisting}
\end{footnotesize}
}

\begin{figure}[!t]
    \centering \includegraphics[width=.99\linewidth]{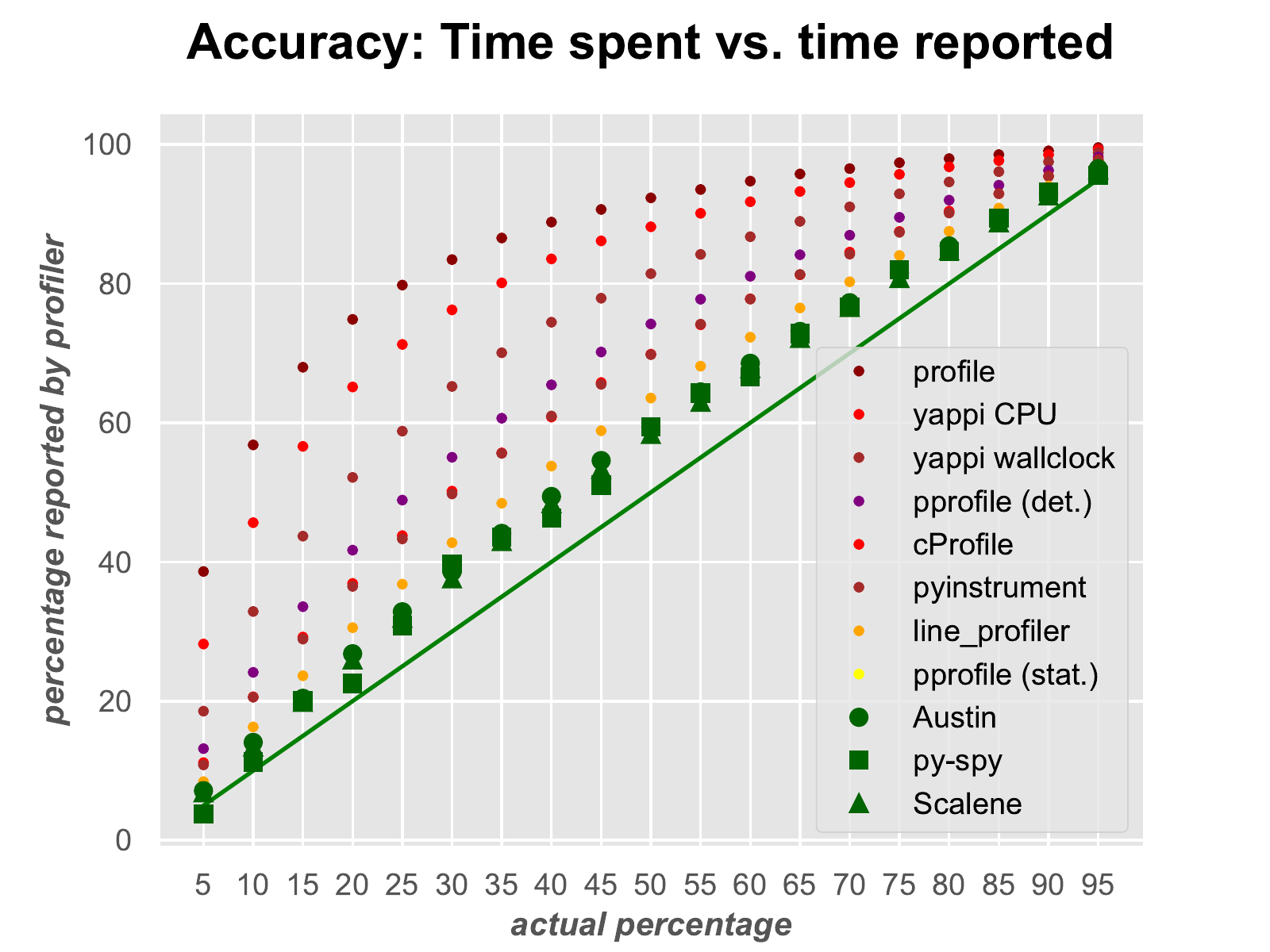}
    \caption{\textbf{CPU Profiling Accuracy: \systemname{} is among the most accurate CPU profilers.} This graph measures the accuracy of profile reports vs. the actual time spent in functions; the ideal is shown by the diagonal line (the amount the profiler reports is exactly the time spent). Some profilers are highly inaccurate ($\S$\ref{sec:cpu-profiling-accuracy}).\label{fig:cpu-accuracy-comparison}}
    \vspace{1em}
\end{figure}

Figure~\ref{fig:cpu-accuracy-comparison} presents the results of this
experiment. The x-axis corresponds to the amount of time
measured while running the variant with a function call (the ground truth), while the
y-axis corresponds to the amount of time reported by each
profiler. The ideal is a diagonal running from the origin.

The results confirm our hypothesis. The trace-based profilers exhibit
a high degree of inaccuracy; trace-based profilers exhibit significant
function bias.  In the worst case, one such profiler reports a
function takes 80\% of execution time while in fact it only consumes
25\%. We conclude that such profilers may be too potentially
misleading to be of practical value for developers.

\paragraph{Summary:} \systemname{} produces accurate profiles on a microbenchmark that stresses function bias, placing it among the most accurate CPU profilers.

\subsection{Memory Profiling Accuracy}
\label{sec:memory-profiling-accuracy}

\begin{figure}[!t]
\centering
\includegraphics[width=\linewidth]{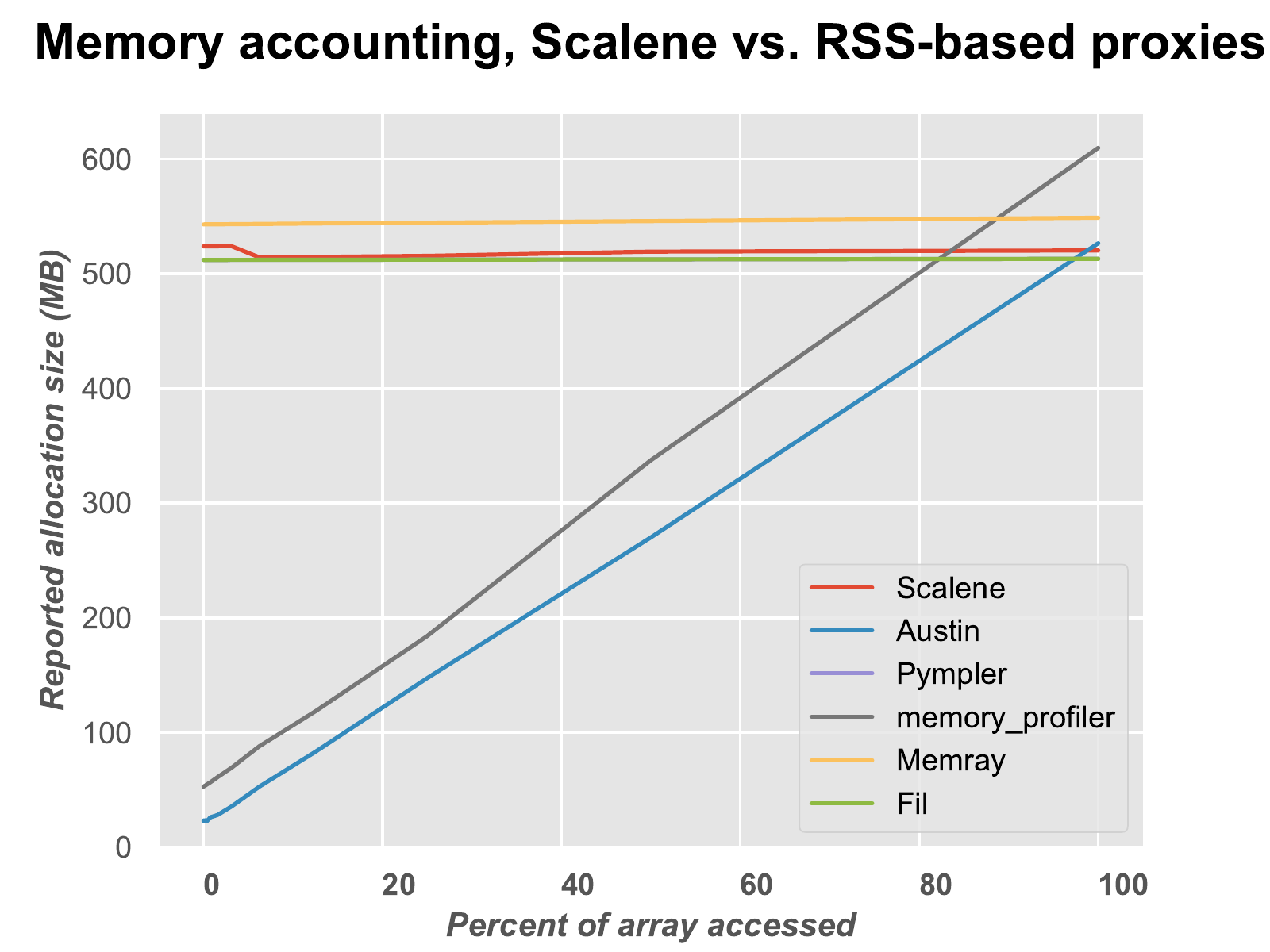}
\caption{\textbf{Memory Profiling Accuracy: \systemname{} produces more accurate memory profiles than resident set size (RSS) based profilers.} Varying the amount of memory accessed causes RSS-based profilers to significantly under-report, and sometimes over-report, the true amount of allocated memory. Interposition-based profilers are far more accurate ($\S$\ref{sec:memory-profiling-accuracy}).\label{fig:rss-bad}}
\end{figure}

We next compare the accuracy of memory profilers with a simple test
designed to explore the effect of using resident-set size instead of
direct memory tracking.  Our hypothesis was that RSS would be an
imperfect proxy, since it corresponds to the use of memory rather than
the allocation of objects. Our test first allocates
a single 512MB array, and then accesses a varying amount of the
array (from 0\% to 100\%).

Figure~\ref{fig:rss-bad} presents the result, confirming our
hypothesis. Both \texttt{memory\_profiler} and
\texttt{Austin} rely on resident set size (RSS) as a proxy for memory consumption.
The figure clearly shows that this can be wildly inaccurate, leading
to under-reporting and even over-reporting the size of the allocated
object. The other profilers directly measure allocation, and produce
much more accurate results. Both \systemname{} and \texttt{Fil} report
within 1\% of the actual size of the allocated object (512MB),
while \texttt{Memray} is within 6\%.

\paragraph{Drawbacks of peak-only profiling:}
\label{sec:drawbacks-peak-only}
Both \texttt{Fil} and \texttt{Memray} only report live objects at the
point of peak memory allocation by a program. This information can be
useful, but it can both exaggerate the potential for reducing memory
and obscure other sources of memory consumption. Consider a program
that allocates and discards a 4GB object, and then allocates a 4GB + 8
byte object. A report that only contains information at the point of
peak allocation will reveal the second object but not the first. That
profile will suggest an enormous opportunity to save memory, but
eliminating the second object entirely would have almost no effect on
peak memory consumption. Unlike peak profilers, \systemname{}
provides information about \emph{all} significant memory allocation
over time, giving programmers a global view of memory consumption.

\paragraph{Summary:} \systemname{}'s memory profiling is highly accurate, while capturing memory consumption over time.

\subsection{CPU Profiling Overhead}
\label{sec:cpu-profiling-overhead}

\begin{table}[!t]
  \begin{small}
\begin{tabular}{lrr}
  \textbf{Benchmark} & \textbf{Repetitions} & \textbf{Time}\\
  \toprule
    $\texttt{async\_tree\_io}_\texttt{none}$ & 22 &{11.9s}\\
    $\texttt{async\_tree\_io}_\texttt{io}$ & 9 &{12.0s}\\
    $\texttt{async\_tree\_io}_\texttt{cpu\_io\_mixed}$ & 14 &{12.3s}\\
    $\texttt{async\_tree\_io}_\texttt{memoization}$ & 16 &{10.6s}\\ 
    \texttt{docutils} & 5 &{12.5s}\\
    \texttt{fannukh} & 3 &{12.1s}\\ 
    \texttt{mdp} & 5 &{13.4s}\\
    \texttt{pprint} & 7 &{12.8s}\\
    \texttt{raytrace} & 25 &{11.1s}\\ 
    \texttt{sympy} & 25 &{11.3s} \\
\end{tabular}
\end{small}
  \caption{\textbf{Benchmark suite:} We conduct our evaluation using the top ten most time consuming benchmarks from the standard \texttt{pyperformance} benchmark suite. For each, we extend their running time by running them in a loop enough times to exceed 10 seconds.\label{fig:benchmark-suite}}
    \vspace{1em}
  
\end{table}

\begin{table}[!t]
    \begin{small} \begin{tabular}{lrrr}
    \textbf{Benchmark} & \textbf{Rate} & \textbf{Threshold} & \textbf{Ratio} \\
    \toprule
    $\texttt{async\_tree\_io}_\texttt{none}$ & 556 & 215 & $3\times$ \\
    $\texttt{async\_tree\_io}_\texttt{io}$   & 524 & 187 & $3\times$ \\
    $\texttt{async\_tree\_io}_\texttt{cpu\_io\_mixed}$ & 719 & 167 & $4\times$\\
    $\texttt{async\_tree\_io}_\texttt{memoization}$ & 375 & 167 & $2\times$\\
    \texttt{docutils}         &  20 & 5 & $4\times$ \\
    \texttt{fannukh}          & 426 & 5 & $85\times$\\
    \texttt{mdp}              & 316 & 6  & $53\times$  \\
    \texttt{pprint}           &  7976 & 23 & $347\times$ \\
    \texttt{raytrace}         & 215 &  7 & $31\times$ \\
    \texttt{sympy}            &  6757 & 10 & $676\times$ \\
    \hline
    \textbf{Median:}    &       &  & \textbf{$18\times$} \\
    \end{tabular}
    \end{small}
    \caption{\textbf{Threshold vs. Rate-Based Sampling:} \systemname{}'s threshold-based sampling tracks footprint with as many as $676\times$ fewer samples than conventional rate-based sampling (median: $18\times$).\label{tab:sampling-comparison}}
    \vspace{1em}
\end{table}

\begin{figure*}[!t]
  \includegraphics[width=\linewidth]{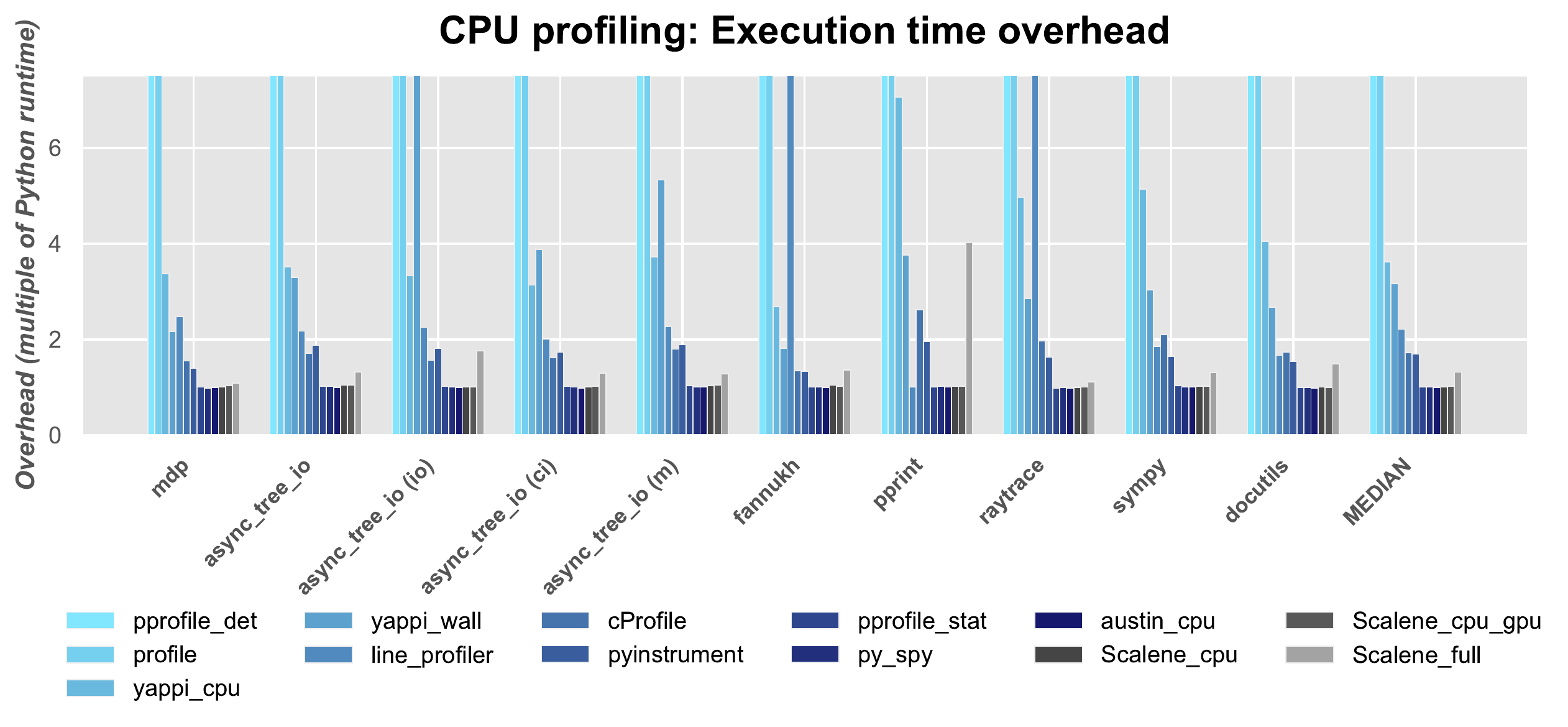}
  \caption{\textbf{CPU
  profiling: \systemname{} has modest overhead.} Despite collecting
  far more detailed information, \systemname{} is competitive with the
  best-of-breed CPU profilers in terms of overhead
  ($\S$\ref{sec:cpu-profiling-overhead}).  The graph
  truncates the slowest profilers; see Table~\ref{fig:benchmark-results} for full data.\label{fig:profiler-overheads}} \vspace{1em}
\end{figure*}

In our evaluation, we use the ten longest-running benchmarks from
\texttt{pyperformance}, the standard suite for evaluating Python performance
(Figure~\ref{fig:benchmark-suite}). We modify these benchmarks to run
in a loop so that they execute for at least 10 seconds on our
experimental platform. We also modify the benchmarks slightly by
adding \texttt{@profile} decorators, as these are required by some
profilers; we also add code to ignore the decorators when they are not
used. Finally, we add a call to \texttt{system.exit(-1)} to
force \texttt{py-spy} to generate
output. Figure~\ref{fig:profiler-overheads} provides the results of running the profilers across
all these benchmarks.

\paragraph{Summary:} In general, \systemname{} imposes low to modest overhead (median:
2\% for CPU+GPU, and 30\% for full functionality), placing it among
the profilers with the lowest overhead.

\subsection{Memory Profiling Overhead}
\label{sec:memory-profiling-overhead}

Next, we evaluate the overhead of memory profilers
(\texttt{memory\_profiler}, \texttt{Fil}, \texttt{Memray}
and \texttt{Austin}), and compare them to \systemname{}. We use the
same benchmarks as we used for measuring runtime overhead for CPU
profilers.

\begin{figure}[!t]
  \includegraphics[width=0.5\textwidth]{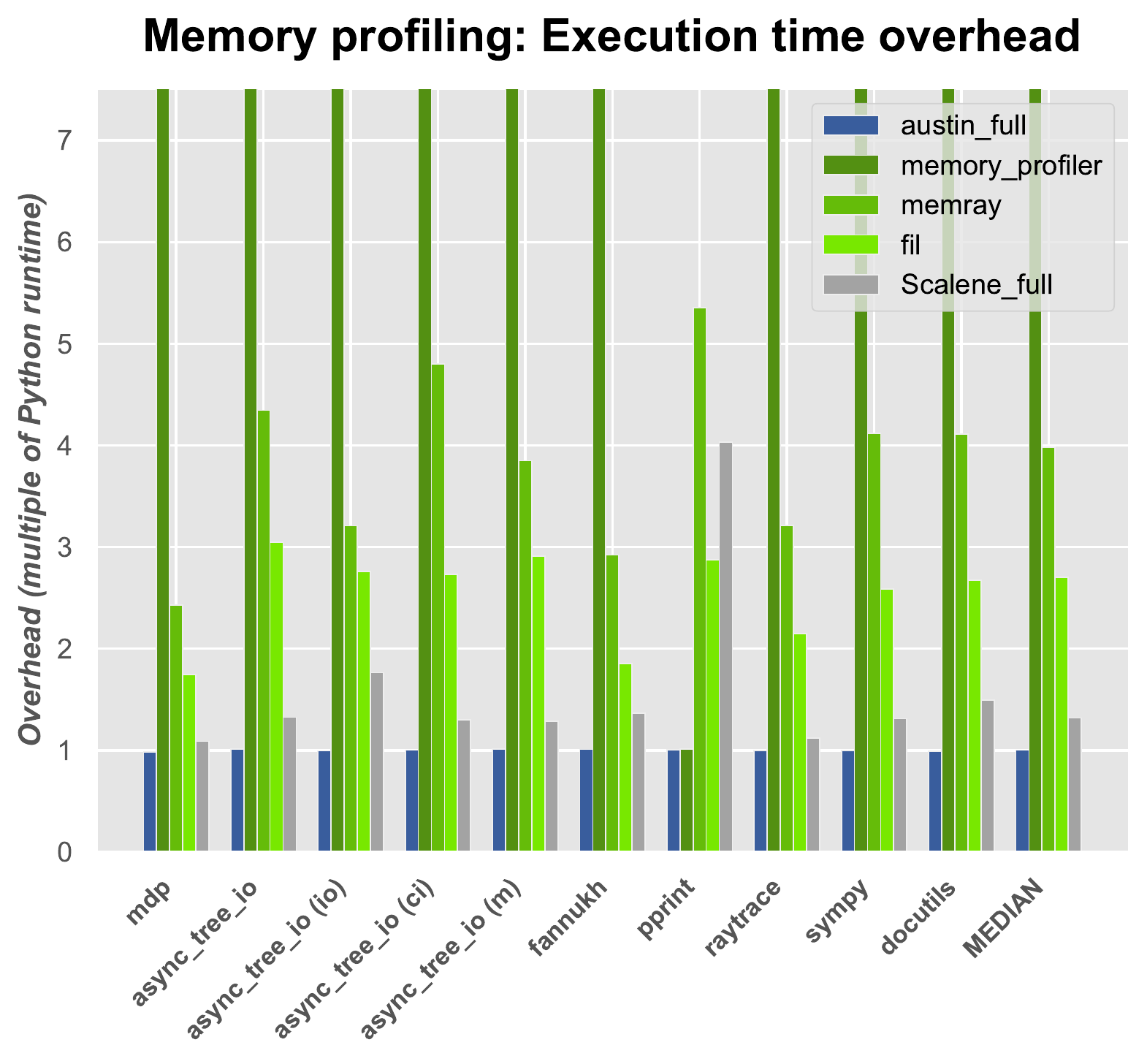} \caption{\textbf{Memory
  profiling overhead: \systemname{} has competitive runtime overhead.}
  Despite collecting far more detailed information, \systemname{} is
  faster than the accurate memory profilers
  ($\S$\ref{sec:memory-profiling-overhead}).\label{fig:memory-profiler-overheads}} \vspace{1em}
\end{figure}

Figure~\ref{fig:memory-profiler-overheads} shows the results. Because it can slow down
execution by at least $150\times$, we omit \texttt{memory\_profiler} from the graph.
\systemname{}'s performance is competitive with the other profilers; while \texttt{Austin}
is faster, as Section~\ref{sec:memory-profiling-accuracy} shows, it
provides inaccurate estimates of memory consumption.

\begin{table*}[!t]
\begin{footnotesize}
\begin{tabular}{lrrrrrrrrrrrrrrrr|r}
& $\textbf{\texttt{mdp}}$ & $\textbf{\texttt{a\_t\_i}}$ & $\textbf{\texttt{(io)}}$ &  $\textbf{\texttt{(ci)}}$ & $\textbf{\texttt{(m)}}$ & $\textbf{\texttt{fannukh}}$ & $\textbf{\texttt{pprint}}$ & $\textbf{\texttt{raytrace}}$ & $\textbf{\texttt{sympy}}$ & $\textbf{\texttt{docutils}}$ & \textbf{Median} \\
\toprule
$\texttt{py\_spy}$ & $0.99\times$ & $1.03\times$ & $1.02\times$ & $1.01\times$ & $1.02\times$ & $1.01\times$ & $1.03\times$ & $1.00\times$ & $1.02\times$ & $1.00\times$ & $1.02\times$ \\
$\texttt{cProfile}$ & $1.55\times$ & $1.71\times$ & $1.57\times$ & $1.62\times$ & $1.80\times$ & $1.35\times$ & $2.63\times$ & $1.98\times$ & $2.11\times$ & $1.74\times$ & $1.73\times$ \\
$\texttt{yappi\_wall}$ & $2.16\times$ & $3.30\times$ & $33.25\times$ & $3.89\times$ & $5.34\times$ & $1.82\times$ & $3.77\times$ & $2.85\times$ & $3.04\times$ & $2.67\times$ & $3.17\times$ \\
$\texttt{yappi\_cpu}$ & $3.38\times$ & $3.52\times$ & $3.33\times$ & $3.14\times$ & $3.72\times$ & $2.69\times$ & $7.07\times$ & $4.97\times$ & $5.14\times$ & $4.05\times$ & $3.62\times$ \\
$\texttt{pprofile\_stat}$ & $1.01\times$ & $1.03\times$ & $1.02\times$ & $1.02\times$ & $1.04\times$ & $1.01\times$ & $1.01\times$ & $0.98\times$ & $1.04\times$ & $1.00\times$ & $1.02\times$ \\
$\texttt{pprofile\_det}$ & $37.80\times$ & $35.06\times$ & $29.30\times$ & $28.09\times$ & $35.85\times$ & $65.19\times$ & $103.73\times$ & $56.23\times$ & $55.68\times$ & $34.78\times$ & $36.83\times$ \\
$\texttt{line\_profiler}$ & $2.48\times$ & $2.18\times$ & $2.25\times$ & $2.01\times$ & $2.27\times$ & $8.92\times$ & $1.01\times$ & $11.59\times$ & $1.86\times$ & $1.67\times$ & $2.21\times$ \\
$\texttt{profile}$ & $14.30\times$ & $14.53\times$ & $13.54\times$ & $12.48\times$ & $15.71\times$ & $10.41\times$ & $55.68\times$ & $20.87\times$ & $26.17\times$ & $15.66\times$ & $15.1\times$ \\
$\texttt{pyinstrument}$ & $1.40\times$ & $1.89\times$ & $1.81\times$ & $1.74\times$ & $1.89\times$ & $1.34\times$ & $1.96\times$ & $1.64\times$ & $1.65\times$ & $1.54\times$ & $1.69\times$ \\
$\texttt{austin\_cpu}$ & $1.00\times$ & $1.01\times$ & $1.00\times$ & $0.99\times$ & $1.02\times$ & $1.00\times$ & $1.01\times$ & $0.99\times$ & $1.02\times$ & $0.99\times$ & $1.00\times$ \\
$\texttt{austin\_full}$ & $0.98\times$ & $1.01\times$ & $0.99\times$ & $1.00\times$ & $1.01\times$ & $1.01\times$ & $1.01\times$ & $1.00\times$ & $1.00\times$ & $0.99\times$ & $1.00\times$ \\
$\texttt{memray}$ & $2.43\times$ & $4.34\times$ & $3.21\times$ & $4.80\times$ & $3.85\times$ & $2.92\times$ & $5.36\times$ & $3.21\times$ & $4.12\times$ & $4.11\times$ & $3.98\times$ \\
$\texttt{fil}$ & $1.75\times$ & $3.05\times$ & $2.76\times$ & $2.73\times$ & $2.91\times$ & $1.85\times$ & $2.88\times$ & $2.15\times$ & $2.58\times$ & $2.68\times$ & $2.71\times$ \\
$\texttt{memory\_profiler}$ & $>150\times$ & $37.90\times$ & $28.42\times$ & $36.32\times$ & $41.90\times$ & $>150\times$ & $1.01\times$ & $>150\times$ & $18.95\times$ & $9.19\times$ & $37.11\times$\\
$\texttt{\systemname{}\_cpu}$ & $1.02\times$ & $1.05\times$ & $1.01\times$ & $1.02\times$ & $1.04\times$ & $1.05\times$ & $1.02\times$ & $1.00\times$ & $1.03\times$ & $1.01\times$ & $1.02\times$ \\
$\texttt{\systemname{}\_cpu\_gpu}$ & $1.03\times$ & $1.05\times$ & $1.02\times$ & $1.02\times$ & $1.05\times$ & $1.02\times$ & $1.03\times$ & $1.01\times$ & $1.03\times$ & $1.01\times$ & $1.02\times$ \\
$\texttt{\systemname{}\_full}$ & $1.09\times$ & $1.33\times$ & $1.76\times$ & $1.30\times$ & $1.28\times$ & $1.36\times$ & $4.03\times$ & $1.12\times$ & $1.31\times$ & $1.49\times$ & $1.32\times$ \\
\end{tabular}
\end{footnotesize}
\caption{\textbf{Detailed profiling overhead (CPU and memory).} All numbers are relative to the Python baseline (no profiling); \texttt{a\_t\_i} refers to the \texttt{async\_tree\_io} benchmark. \label{fig:benchmark-results}}
\end{table*}

\paragraph{Log file growth:}
Some memory profilers feature a surprising other source of overhead.
Two of the memory profilers, \texttt{Memray} and \texttt{Austin},
produce detailed (and copious) logs of memory activity that may limit
their usefulness for profiling long-lived applications.

\texttt{Memray} deterministically logs information including all allocations, all updates to the Python
stack, and context switches, which it later post-processes for
reporting. \texttt{Austin} similarly generates logs meant to be
consumed by an external tool. These files can grow rapidly: in our
tests, \texttt{Memray}'s output file grows by roughly 3MB/second,
while \texttt{Austin}'s grows by 2MB/second.

By contrast, \systemname{} only records samples when memory
consumption grows or shrinks by a large amount
($\S$\ref{sec:memory-sampling}), leading to vastly smaller logs. For
example, when running the \texttt{mdp} benchmark,
\texttt{Austin}'s log file consumes 27MB and \texttt{Memray}'s log file consumes
almost 100MB, while \systemname{}'s log file consumes just 32K.

\paragraph{Summary:} Among the accurate memory profilers, \systemname{} operates with the lowest overhead (median: $1.32\times$ vs. $3.98\times$ (\texttt{memray}) and $2.71\times$ (\texttt{Fil}), while capturing memory usage over time and producing small log files.

  

  \section{Case Studies}
  \label{sec:case-studies}

This section includes reports on real-world experience by external
developers using \systemname{} to identify and resolve performance
issues. For each, we identify the features of \systemname{} that
were instrumental in enabling these optimizations.

\paragraph{Rich:}
A user reported severe slowness when printing large tables to the
developer of Rich~\cite{rich}, an immensely popular Python library for
formatting text in the terminal (downloaded over 130 million times, with
41K stars on GitHub). When Rich's developer profiled it
using \systemname{}, he identified two lines occupying a
disproportional amount of runtime. \systemname{} indicated that a call
to \texttt{isinstance} was taking an unexpectedly large amount of
time--though each call takes very little time, the developer reported
that it was being called 80,000 times. Rich's developer replaced these
calls with a lower-cost function, \texttt{hasattr}. In our
benchmarks, \texttt{isinstance} (when marked as a runtime protocol
via \texttt{@typing.runtime\_checkable}) can run over $20\times$
slower than \texttt{hasattr}. The developer also indicated that an
unnecessary copy was being performed once every cell. Optimizing these
calls led to a reported 45\% improvement in runtime when rendering a
large table. \textbf{{[Feature: Fine-grained CPU profiling, copy volume.]}}

\paragraph{Pandas -- Chained Indexing:}

A developer was seeing suboptimal performance in their code using
Pandas~\cite{reback2020pandas}. \systemname{} identified that a list
comprehension performing nested indexes into a Pandas dataframe was
taking an unexpectedly large amount of time and resulting in a
significant amount of copy volume. The developer noted that the first
level of indexing was repeatedly using a string that was loop
invariant; the way this was being done in Pandas caused it to perform
copies rather than using views, a problem known as {chained
indexing} (\url{https://pandas.pydata.org/pandas-docs/stable/user_guide/indexing.html\#returning-a-view-versus-a-copy}). After
manually hoisting this outer indexing operation, the developer
obtained an $18\times$ speedup. \textbf{{[Features: Copy volume and fine-grained CPU profiling.]}}

\paragraph{Pandas -- \texttt{concat} and \texttt{groupby} queries:}

An instructor had their students use \systemname{} in a tutorial
designed to teach higher performance Pandas. The instructor found
that \systemname{} revealed significant issues in both
performance and space consumption when using Pandas. 
First, \systemname{} revealed that calling \texttt{concat} on Pandas
dataframes was using more memory than anticipated. \systemname{}'s copy volume reporting
revealed that the problem was that \texttt{concat}
copies all the data by
default (\url{https://pandas.pydata.org/pandas-docs/stable/reference/api/pandas.concat.html\#pandas.concat}),
effectively doubling memory usage when managing large
dataframes. Second, \systemname{} confirmed that excessive RAM usage in some
\texttt{groupby} operations is due to copying of the groups;
this bug has been reported to the Pandas developers
(\url{https://github.com/pandas-dev/pandas/issues/37139}).
Restructuring the \texttt{groupby} operation
reduced memory consumption by a further 1.6GB.
\textbf{{[Features: Fine-grained CPU and memory profiling, copy volume.]}}

\paragraph{NumPy vectorization:}

A graduate student was using NumPy to implement classification with
gradient descent and was seeing extremely low
performance. \systemname{} showed that 99\% of the time was being
spent in Python (rather than native code), indicating that his code
was not vectorized. In other words, the code was not expressed in a
way that allowed NumPy to efficiently compute vector operations (using
native code). Guided by \systemname{}'s feedback, the graduate student
gradually improved the performance from 80 iterations per minute to
10,000 per minute, a $125\times$ improvement.
\textbf{{[Feature: Fine-grained native vs. Python CPU profiling.]}}

\paragraph{Semantic Scholar:}

Semantic Scholar reports that they have been using \systemname{} as
part of their tool suite for operationalizing their machine learning
models. Recently, they found that a model was cost-prohibitive and put
an entire product direction in jeopardy. They generated a set of test
data and ran their models with \systemname{}. \systemname{}'s output
was able to pinpoint the issues and help them validate that their
changes were having an impact. While iteratively using
\systemname{} while applying optimizations, they were ultimately
able to reduce costs by 92\%. Additionally, \systemname{} allowed
Semantic Scholar's developers to quickly determine what fraction of
their runtime would benefit from hardware acceleration and what
CPU-bound code they needed to optimize in order to achieve their
goals.
\textbf{{[Features: Simultaneous, fine-grained CPU, memory, and GPU profiling.]}}

\paragraph{Summary:}

In nearly all of the cases described above, \systemname{} was either
invaluable or provided additional help that narrowed down performance
issues, including several unique features of \systemname{}: separation
of native from Python time, copy volume, GPU profiling, and its
ability to simultaneously measure memory and CPU usage. Though other
tools can separately identify high RAM usage or slow code, past tools
would either misattribute the location of usage due to the use of
resident set size as a metric (unlike \systemname{}'s accurate memory
profiling approach) or not be able to simultaneously measure memory
usage and CPU usage. The insights generated by \systemname{} were
actionable, yielding substantial improvements in execution time and
space, as well as cost reduction.

  \section{Related Work}
  \label{sec:related_work}

There is an extensive history of profilers; we focus our
attention here on profilers that specifically support Python.
The Python ecosystem has given rise to a proliferation of Python
profilers, most of which have not been discussed in the academic
literature. This section describes the most prominent profilers;
Figure~\ref{fig:profiler-comparison} provides a diagrammatic overview.


We first survey CPU-only profilers.
We divide them into two categories: \emph{deterministic}
(tracing-based) ($\S$\ref{sec:deterministic-cpu}) and \emph{sampling-based} ($\S$\ref{sec:sampling-based-cpu}).
We then discuss memory profilers ($\S$\ref{sec:memory-profilers}), ML-specific profilers ($\S$\ref{sec:profilers-ml}),
other Python profilers ($\S$\ref{sec:other-profilers}), and general profilers with Python support ($\S$\ref{sec:profilers-python-support}), and touch on more distantly related profilers for other languages ($\S$\ref{sec:non-python-profilers}).

\subsection{Deterministic CPU profilers}
\label{sec:deterministic-cpu}

Python provides built-in tracing support (\texttt{sys.settrace}) that
several profilers build upon.  The tracing facility, when activated,
triggers a callback in response to a variety of events, including
function calls and execution of each line of code. This deterministic,
instrumentation-based approach leads to significant inaccuracies due
to its probe effect, as Section~\ref{sec:cpu-profiling-accuracy}
shows. Because of the overhead of tracing, they are also the slowest
profilers.

\paragraph{Function-granularity:}
Python includes two built-in function-granularity profilers,
\texttt{profile}~\cite{profile} and
\texttt{cProfile}~\cite{cprofile}. The primary difference between
these two profilers is that \texttt{cProfile}'s callback function is
implemented in C, making it much faster ($1.7\times$ slowdown
vs. $15.1\times$) and somewhat more accurate than \texttt{profile}.
Another profiler, \texttt{yappi}, operates in two modes, wall clock
time (sample-based) and CPU time (deterministic); it is among the most
inaccurate of CPU profilers, with slowdowns ranging from $1.8\times$ to
$33.3\times$.

\paragraph{Line-granularity:}
\texttt{pprofile}~\cite{pprofile} comes in two flavors: a deterministic and a ``statistical'' (sampling-based) profiler.
Both flavors correctly work for multithreaded Python programs,
unlike \texttt{line\_profiler}~\cite{line-profiler}. All of these
report information at a line granularity. \texttt{pprofile\_det} 
imposes a median overhead of $36.8\times$,
while \texttt{line\_profiler}'s median overhead is $2.2\times$.

\subsection{Sampling-based CPU profilers}
\label{sec:sampling-based-cpu}

Sampling-based profilers are both more efficient and often more accurate
than the deterministic profilers.  These
include \texttt{pprofile\_stat},
\texttt{py-spy}~\cite{py_spy},
and \texttt{pyinstrument}~\cite{pyinstrument}. Their overhead is
between $1\times$ and $1.7\times$, comparable to \systemname{}.
\texttt{pprofile\_stat} incorrectly ascribes zero runtime to
execution of native code or code in child threads ($\S$\ref{sec:cpu-implementation}).

Compared to past CPU-only profilers, \systemname{} is nearly as
fast or faster, more accurate, and provides more detailed CPU-related
information, breaking down time spent into Python, native, or system
time.

\subsection{Memory profilers}
\label{sec:memory-profilers}

\texttt{memory\_profiler} is a deterministic memory profiler that uses 
Python's trace facility to trigger it after every line of
execution~\cite{MemoryProfiler}. By default, it measures the RSS after
each line executes and records the change from the previous
line. \texttt{memory\_profiler} also does not
support Python applications using threads or multiprocessing.

\texttt{Fil} measures the peak allocation of the profiled program by
interposing on system allocator functions and forcing Python to use
the system allocator (instead of Python's \texttt{Pymalloc})~\cite{fil-profiler}.
\texttt{Fil} records a full stack
trace whenever the current memory footprint exceeds a previous maximum.
On exit, it produces a flamegraph~\cite{10.1145/2909476} of call
stacks responsible for memory allocation at the point of maximum
memory consumption. The Fil website reports that it supports
threads (``In general, Fil will track allocations in threads correctly.''~\cite{fil-threading}).
However, in
our tests, Fil (version 2022.6.0) fails to ascribe any memory
allocations to threads. Fil also does not currently support
multiprocessing.

\texttt{Memray} is a recently released (April 2022), Linux-only memory profiler
that deterministically tracks allocations and other profiler
events~\cite{memray-profiler}. \texttt{Memray}
interposes upon the C allocation functions and optionally on
the \texttt{pymem} functions, letting it distinguish native from
Python allocations.

The only previous CPU+memory profiler we are aware of
besides \systemname{} is \texttt{Austin}~\cite{austin}.
\texttt{Austin} samples the frames of all running threads at a specified time interval.
Austin profiles from its own process outside of
the program being profiled, reducing its performance overhead.

\subsection{Profilers for Machine Learning Libraries}
\label{sec:profilers-ml}

Two widely used machine learning libraries, TensorFlow and
PyTorch~\cite{paszke2019pytorch}, include their own
profilers~\cite{tf-profiler,pytorch-profiler}. Both profilers are
targeted at identifying performance issues specific to deep learning
training and inference. For example, the PyTorch profiler can
attribute runtime to individual operators (running inside PyTorch's
native code).

NVIDIA's Deep Learning profiler (\texttt{DLprof})~\cite{dlprof}
provides similar functionality for either PyTorch or
TensorFlow. Unlike \systemname{}, these profilers are specific to
machine learning workloads and are not suitable for profiling
arbitrary Python code. These profilers are complementary
to \systemname{}, which aims to be a general-purpose profiler. They
also lack many of \systemname{}'s features.

\subsection{Other Python Profilers}
\label{sec:other-profilers}

PieProf aims to identify and surface specific types of inefficient
interactions between Python and native
code~\cite{DBLP:conf/sigsoft/Tan0LRSS021}. PieProf leverages data
gathered from on-chip performance monitoring units and debug registers
combined with data from \texttt{libunwind} and the Python interpreter
to identify redundant loads and stores initiated by user-controlled
code. It surfaces pairs of redundant loads and stores for the
developer to potentially optimize. PieProf is not publicly available,
so it was not possible to compare it to \systemname{}; we view its
functionality it as orthogonal and complementary.

\subsection{Profilers with Python Support}
\label{sec:profilers-python-support}

Several non-Python specific conventional profilers offer limited
support for Python.  Intel's VTune profiler~\cite{vtune} can attribute
its metrics to Python lines, with a number of caveats, including ``if
your application has very low stack depth, which includes called
functions and imported modules, the VTune Profiler does not collect
Python data.''~\cite{vtune-user-guide}. VTune does not directly
distinguish between time spent in Python code and time spent in native
code and does not track Python memory allocations. Google Cloud
Profiler~\cite{google-cloud-profiler} only profiles Python execution
time, but neither distinguishes between Python and native time nor
does it perform memory profiling for Python. Both lack most of \systemname{}'s
other features.

Python 3.12, the current development version of Python, recently
(November 2022) added support for use with the \texttt{perf} profiler
on Linux platforms by reporting function names in
traces~\cite{python-perf}. Using \texttt{perf} in this mode only
measures performance counters or execution time. Unlike \systemname{},
\texttt{perf} does not measure memory allocation, or attribute runtime
(Python or native) to individual lines of Python code.


\subsection{Non-Python Profilers}
\label{sec:non-python-profilers}

\texttt{AsyncProfiler} is a Java profiler
that, like \systemname{}, profiles both CPU and
memory~\cite{asyncprofiler}. \texttt{AsyncProfiler} is a sampling
profiler that avoids the \emph{safepoint bias
problem}~\cite{DBLP:conf/pldi/MytkowiczDHS10}. Since Python does not
have safepoints (all garbage collection happens while the global
interpreter lock is held), Python profilers cannot suffer from this
bias. Instead, as we show, they can suffer from function bias
($\S$\ref{sec:cpu-profiling-accuracy}).
Similarly, \texttt{pprof} is a profiler for the Go language that can
report both CPU and memory~\cite{pprof}. Both profilers use rate-based
memory sampling ($\S$\ref{sec:memory-sampling}).

  \section{Conclusion}
  \label{sec:conclusion}

This paper presents \systemname{}, a novel profiler for
Python. \systemname{} both sidesteps and exploits characteristics of
Python to deliver more actionable information than past
profilers, all with high accuracy and low overhead. Its suite of novel
algorithms enables \systemname{}'s holistic reporting of Python
execution. \systemname{} has been released as open source
at \url{https://github.com/plasma-umass/scalene}.

\section*{Acknowledgements}

We thank \systemname{}'s users for their feature requests, questions,
and bug reports, which have helped shape and guide this research. We
are most grateful to users who contributed pull requests or worked
with us to resolve compatibility issues, including Raphael Cohen,
James Garity, Ryan Grout, Friday James, and Marguerite Leang. We thank
the users who contributed their experiences, reported here as case
studies: Will McGugan, Ian Oszvald, Donald Pinckney, Nicolas van
Kempen, and Chris Wilhelm. Finally, we thank the reviewers of this
paper, whose feedback helped improve not only the paper but
also \systemname{} itself.

This material is based upon work supported by the
National Science Foundation under Grant No. 1954830.

  {
  \bibliographystyle{abbrv}
  \bibliography{emery,scalene}
  }
  
\end{document}